# First Orbital Solution and Evolutionary State for the Newly Discovered Eclipsing Binaries USNO-B1.0 1091-0130715 and GSC-03449-0680

M. M. Elkhateeb[1, 2], M. I. Nouh[1, 2] and R. H. Nelson

[1]Astronomy Department, National Research Institute of Astronomy and Geophysics, 11421 Helwan, Cairo, Egypt

E-mail: abdo_nouh@hotmail.com, Fax: +202 2554 8020

[2]Physics department, College of Science, Northern Border University, 1321 Arar, Saudi Arabia

[3]1393 Garvin Street, Prince George, BC, Canada, V2M 3Z1

**Abstract:** First photometric study for the newly discovered systems USNO-B1.0 1091-0130715 and GSC-03449-0680 were applied by means of recent windows interface version of the Wilson and Devinney code based on model atmospheres by Kurucz (1993). The accepted models reveal some absolute parameters for both systems, which used in adopting the spectral type of the systems components and follow their evolutionary status. Distances to each systems are and physical properties are estimated. Comparisons of the computed physical parameters with stellar models are discussed.

## 1. Introduction

Study of the eclipsing binary systems gives information about their stellar properties, while physical parameters estimated from the light curve modeling of these systems could be used to follow their evolutionary state. The systems USNO-B1.0 1091-0130715 and GSC-03449-0680 are two newly discovered eclipsing binaries, their coordinates are listed in Table (1). To the best of our knowledge, no further light curve analysis was performed for both of them. The present paper is a continuation of a program started to study some newly discovered eclipsing binaries, Elkhateeb et al. (2014a, b and c).

The structure of the paper is as follows: section 2 displays the basic information about the studied systems and light minima derived from their observations. Section 3 devoted to the light curve modeling. In section 4 we discuss the evolutionary status of both systems. Summary of the results and conclusion are outlined in section 5.

## 2. Times of Minima and Period Behavior

### 2.1. USNO-B1.0 1091-0130715

The system USNO-B1.0 1091-0130715 (P = $0^d.5997$) was observed and reported as a variable star by Koff (2004) and listed as eclipsing binary of beta Lyr type (semi-detached) in SIMBAD data base. The system was observed in the period from January 25 to February 21, 2003 using a Meade LX-200 0.25-m f/10 SCT telescope of Antelope Hills Observatory with Apogee AP-47 1K X 1K CCD in V filter. Figure (1) displays the observed light curves of the system USNO-B1.0 1091-0130715 in V filter. Two times of minima were estimated from the observed light curve by means of the Minima V2.3 package (Nelson 2006) based on the Kwee and Van Worden (1956) fitting method. The new minima appear in Table (2) together with the Epoch (E), and (O-C)'s calculated using the first ephemeris by Koff (2004):

$$\text{Min I} = 2452678.7571 + 0.5997 * E \tag{1}$$

The (O-C) values are insufficient to conclude any trend about the possible variation of the period.

### 2.2 GSC-03449-0680

The variability of the system GSC-03449-0680 was discovered by Krajci et al. (2005). The system was classified as a detached system (EA) with a period of P = 0.560644 day. First observations for the system were carried out by Krajci et al. (2005) in V and R pass band using a 0.33 m reflector telescope of Sylvester robotic observatory, which attached to SBIG ST-70 CCD camera. Observations of the system were repeated by Nelson (2005), Krajci (2006, 2007). Figure (2) displays the observed light curves of the system GSC-03449-0680 in V and R filter. A total of nine new minima (five primary and 4 secondary) were published, and the corresponding residuals (O-C) were calculated using the first linear ephemeris by Krajci et al (2005) (Eq. 2), and listed in Table (2).

$$\text{Min I} = 2453404.8768 + 0.560644 * E \tag{2}$$

The general trend of the (O-C) values can be represented by a linear fit, Figure (3), which shows a period decrease by 0.17 second than Krajci et al. (2005). The equation represented the fit could be written as:

$$\text{Min I} = 2453404.87677 + 0.560642 * E \qquad (3)$$

## 3. Photometric Analysis

Photometric analysis for the systems USNO-B1.0 1091-0130715 and GSC-03449-0680 were carried out using (2009) version of the Wilson and Devinney code (Nelson, 2009), which based on model atmospheres by Kurucz (1993). The initial value for the temperature of the primary star ($T_1$) was estimated using the (J-H) infrared color index for each system listed in SIMBAD (http://simbad.u-strasbg.fr/simbad/). Then we used (J-H) color index temperature relation by Tokunaga (2000) to estimate the corresponding temperature for each color index. All individual observations of the observed light curves in each band were analyzed instead of normal points because the normal light curves do not reveal the true light variation of the system. We adopted gravity darkening and bolometric albedo exponents for the convective envelopes ($T_{eff}$ < 7500 K), where $A_1 = A_2 = 0.5$ (Rucinski 1969) and $g_1 = g_2 = 0.32$ (Lucy, 1967). Bolometric limb darkening were adopting using Tables of Van Hamme (1993) based on the logarithmic law for the extinction coefficients. Through the light curve solution, the commonly adjustable parameters employed are; the orbital inclination (i), the mass ratio (q); temperature of the primary component ($T_1$) and secondary one ($T_2$), surface potentials of the stars ($\Omega_1$, $\Omega_2$) and the monochromatic luminosity of primary star ($L_1$). Relative brightness of secondary star was calculated by the stellar atmosphere model.

### 3.1. USNO-B1.0 1091-0130715

The light curve of the system USNO-B1.0 1091-0130715 in V band was analyzed using W-D code (Nelson, 2009) under the condition of Mode 4 (semidetached). Trails were made to estimate set of parameters which represented the observed light curves. Best photometric fitting was reached after several runs, which shows that the primary component is more massive and hotter than the secondary one, with a temperature difference of about 2291 K. The accepted parameters are listed in Table (3), while Figure (4) displays the observed light curves together with the synthetic curves in V pass band. According to the accepted orbital solution, the components of the system USNO-B1.0 1091-0130715 are of spectral type A6 and G3 respectively (Popper, 1980).

Using the software package Binary Maker 3.03 (Bradstreet and Steelman, 2004), a three dimensional geometrical structure based on the calculated parameters of the system is displayed in Figure (5).

### 3.2. GSC-03449-0680

A photometric study for the observed light curves of the system GSC-03449-0680 was applied in V and R pass band. During the calculations, Mode 5 (semidetached – Algole) of W-D code was applied. The accepted solution shows that the primary component belong to spectral type of F6 and the secondary is K8 with temperature difference $\Delta T \sim 2336$ K. Table (3) listed the parameters of the accepted model, while Figure (6) displays the reflected observed points in V and R pass band together with the corresponding theoretical light curves. Figure (8) displays the corresponding three dimensional structure of the system.

### 4. Evolutionary State of the Systems

Spectroscopic observations are one of the important sources for physical parameters estimation. Because of the studied systems are completely newly discovered systems and there is no available radial velocity for both of them, we estimated the absolute physical parameters for the systems components by means of the empirical $T_{eff}$ – Mass relation of Harmanac (1988). The estimated physical parameters show that the primary components in both systems are massive than the secondary one. The systems photometric and absolute properties were used to calculate the distance (d) of both systems (d = $10^{(m - Mv + 5)/5}$), where **m** and **Mv** are the apparent and absolute magnitudes respectively. The results reveal that the average distance for the system USNO-B1.0 1091-0130715 is 2239 pc, while equal to 1308 pc for the system GSC-03449-0680.

We used the physical parameters listed in Table (4) to investigate the current evolutionary state of the two systems. In Figures (8) and (9) we plotted the components of the two systems on the mass–luminosity (M-L) and mass–radius (M-R) relations along with the evolutionary tracks computed by Girardi et al. (2000) for both zero age main sequence stars (ZAMS) and terminal age main sequence stars (TAMS) with metalicity $z = 0.019$.

As it is clear from the figures, the components of the two systems are located near the ZAMS for M-R relation; therefore the components of the two systems appear to be main sequence stars.

For the M-L relation, the components of the two systems are on the ZAMS track, except the secondary component of the system USNO-B1.0 1091-0130715 which shows that the star have lower luminosity than that expected from that ZAMS stars.

To test the matching between the masses arose from the orbital solution and that for the stellar models we locate the components of the two systems on the $T_{eff}$-luminosity relation for singles stars, we used for this purpose, the non-rotated evolutionary models of Ekström et al. (2012) in the range 0.8–120 M⊙ at solar metalicity (z = 0.014). In Figure (10) we used the tracks for the masses appropriate that of the orbital solution. Both components of GSC-03449-00680 display fair fit. The primary component of USNO-B1.0 1091-0130715 reveals fair fit but the secondary one has the same behavior of the M-L relations.

The mass-effective temperature relation (M–$T_{eff}$) relation for intermediate and low mass stars (Malkov, 2007) is displayed in Figure (11). The locations of the masses and radii of the two systems on the diagrams revealed a good fit for the components of GSC-03449-00680. Comparison of the masses and radii of USNO-B1.0 1091-0130715 provides a good fit for the primary component and poor fit for the secondary component. This gave the same behavior of the system on the M-L relation.

The results presented here are a preliminary one, further spectroscopic and photometric observations are needed to obtain good models for the two systems.

## 5. Discussion and Conclusion

New semidetached system USNO-B1.0 1091-0130715 and detached system GSC-03449-0680 were discovered in 2004 and 2005 respectively. A complete light curves were observed for both systems and new times of minima were calculated. The first O-C diagram for the system GSC-03449-0680 was established. Results of the first photometric analysis for both systems showed that the primary components in both systems are hotter and massive than the secondary one. Distance to each system was derived and their components spectral type was adopted.

Evolutionary state of the systems under study has been investigated to explore their behavior on the M-R and M-L relations.

Table (1): Coordinates of the studied systems with their comparison and the check stars

| Star Name | RA (2000.0) | Dec (2000.0) |
|---|---|---|
| Variable (USNO-B1.0 1091-0130715) | 06 h 53' 23.5" | 19 h 10' 24.8" |
| Comparison (GSC-03449-0707) | | |
| Variable (GSC-03449-0680) | 10 h 45' 54.6" | 52 h 16' 26.4" |
| Comparison (GSC-03449-0707) | 10 h 45' 30.8" | 52 h 10' 14.18" |

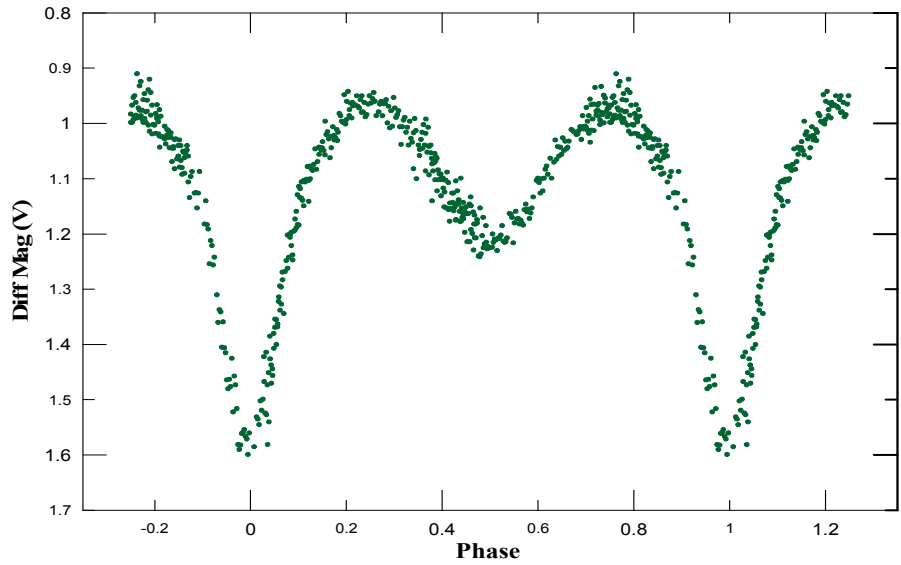

Figure (1): Observed light curves of the system USNO-B1.0 1091-0130715 in V filter.

Table (2): Published times of minima for the systems USNO-B1.0 1091-0130715 and GSC-03449-0680.

| Star | HJD | Error | Type | E | (O-C) | Refrence |
|---|---|---|---|---|---|---|
| USNO-B1.0 1091-0130715 | 2452669.7667 | 0.0018 | I | -15 | 0.0051 | This Paper |
|  | 2452691.6438 | 0.0006 | II | 21.5 | -0.0069 | This Paper |
| GSC-03449-0680 | 2453058.6800 | 0.0010 | II | -617.5 | 0.0009 | Nelson (2005) |
|  | 2453058.9600 | 0.0010 | I | -617 | 0.0006 | Nelson (2005) |
|  | 2453066.8090 | 0.0010 | I | -603 | 0.0005 | Nelson (2005) |
|  | 2453074.9380 | 0.0010 | II | -588.5 | 0.0002 | Nelson (2005) |
|  | 2453077.7440 | 0.0010 | II | -583.5 | 0.0030 | Nelson (2005) |
|  | 2453404.8768 | 0.0003 | I | 0.00 | 0.0000 | Krajci (2006) |
|  | 2454075.9657 | 0.0002 | I | 1197 | -0.0020 | Krajci (2007) |
|  | 2454174.6384 | 0.0002 | I | 1373 | -0.0026 | Krajci (2007) |

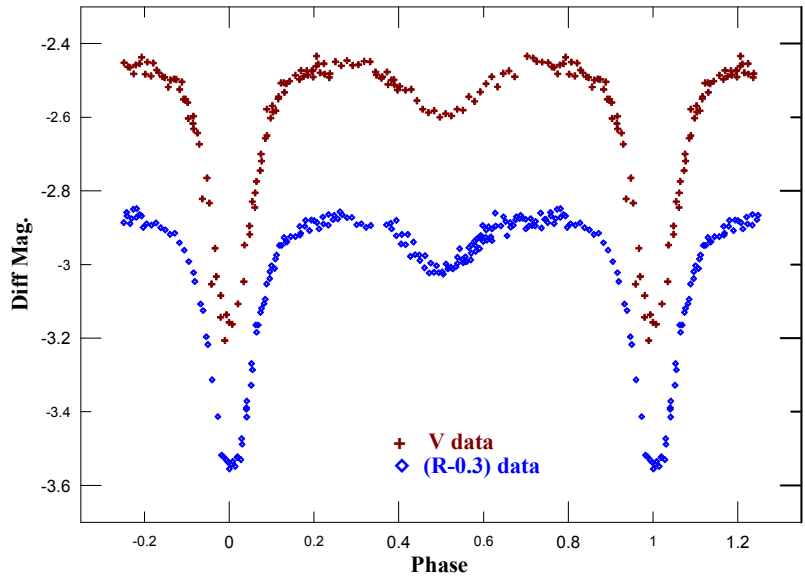

Figure (2): Observed light curves of the system GSC-03449-0680 in VR filters.

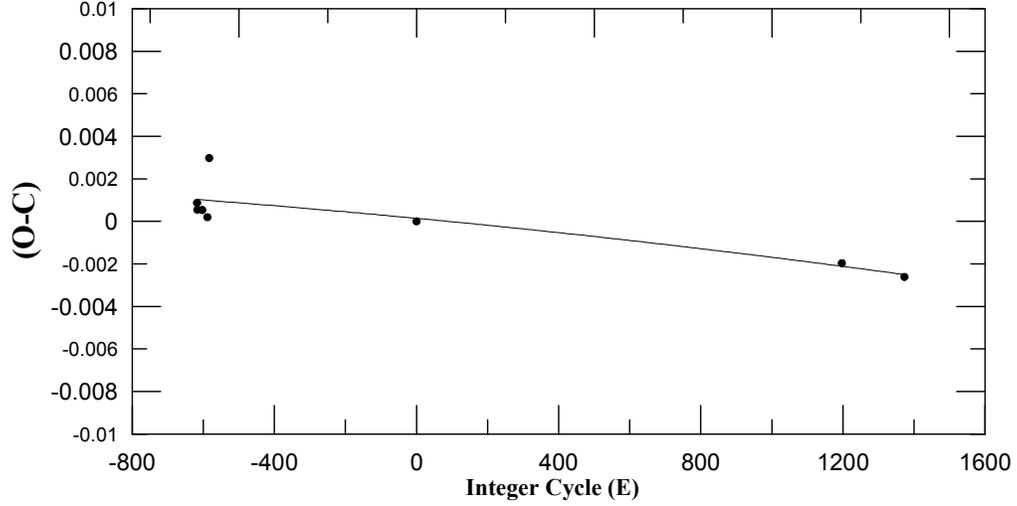

Figure (3): Period behavior of the system GSC-03449-0680.

Table (3): Photometric solution for USNO-B1.0 1091-0130715 and GSC-03449-00680.

| Parameter | USNO-B1.0 1091-0130715 | GSC-03449- 00680 |
| --- | --- | --- |
| $i\ (^0)$ | 84.46±0.62 | 73.10±0.11 |
| $g_1 = g_2$ | 0.5 | 0.5 |
| $A_1 = A_2$ | 0.32 | 0.32 |
| $q = M_2 / M_1$ | 0.6496±0.0052 | 0.7420±0.0069 |
| $\Omega_1$ | 3.1536 | 3.6260±0.0166 |
| $\Omega_2$ | 4.0188±0.0232 | 3.3168 |
| $\Omega_{in}$ | 3.1536 | 3.3168 |
| $\Omega_{out}$ | 2.7771 | 2.8945 |
| $T_1\ (^0K)$ | 8004±36 | 6449±23 |
| $T_2\ (^0K)$ | 5713±25 | 4113±12 |
| $r_1$ pole | 0.3923±0.0020 | 0.3438±0.0008 |
| $r_1$ side | 0.4147±0.0023 | 0.3571±0.0010 |
| $r_1$ back | 0.4439±0.0022 | 0.3745±0.0014 |
| $r_2$ pole | 0.2265±0.0071 | 0.3314±0.0008 |
| $r_2$ side | 0.2299±0.0076 | 0.3470±0.0009 |
| $r_2$ back | 0.2360±0.0085 | 0.3788±0.0008 |
| $\sum (O-C)^2$ | 0.01744 | 0.02699 |

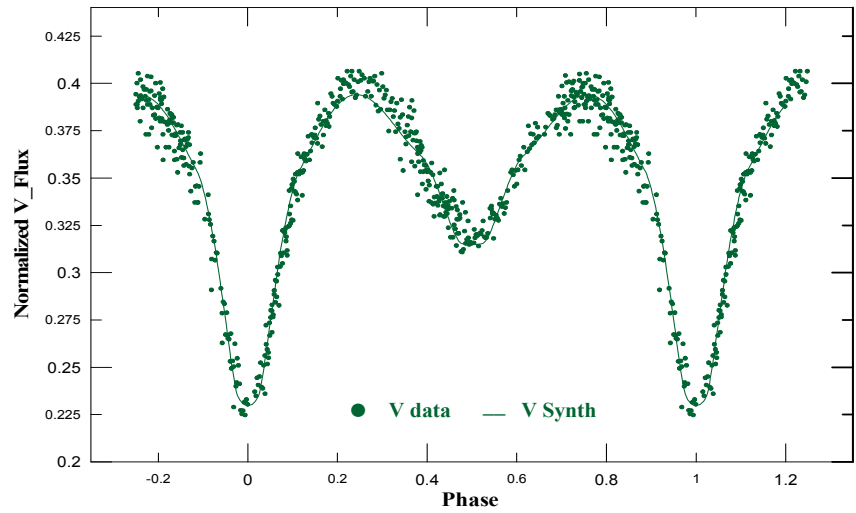

Figure (4): Observed and synthetic light curves for the system USNO-B1.0 1091-0130715 in V filters.

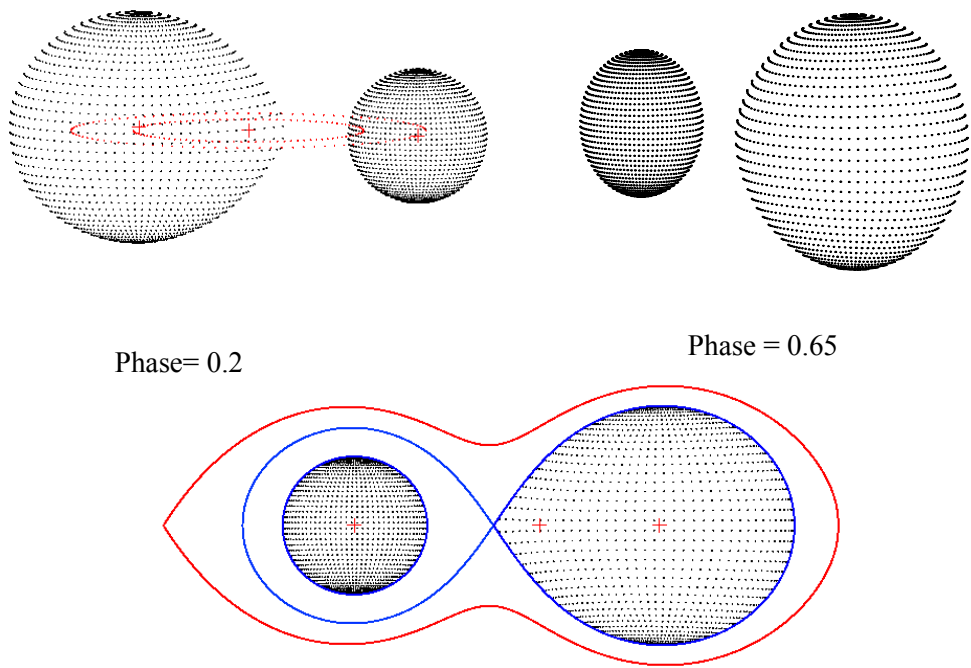

Phase= 0.2

Phase = 0.65

Figure (5) : Geometric structure of the binary system USNO-B1.0 1091-0130715

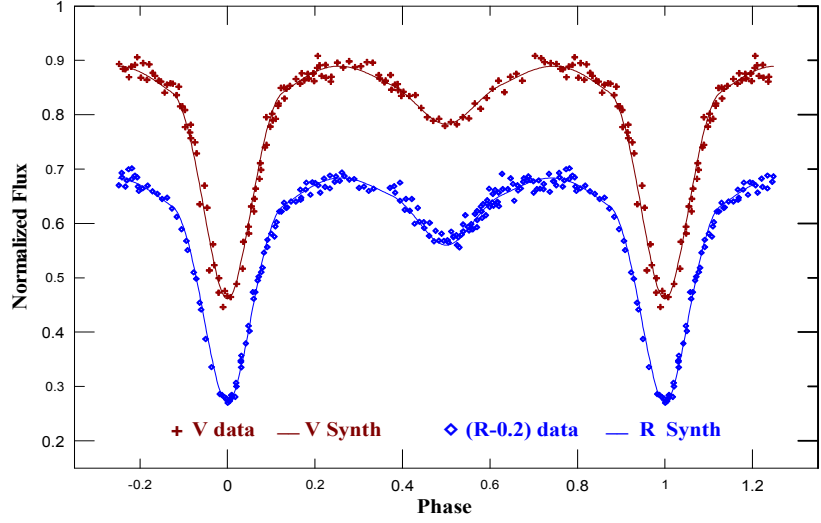

Figure (6): Observed and synthetic light curves for the system GSC-03449-0680 in VI filters.

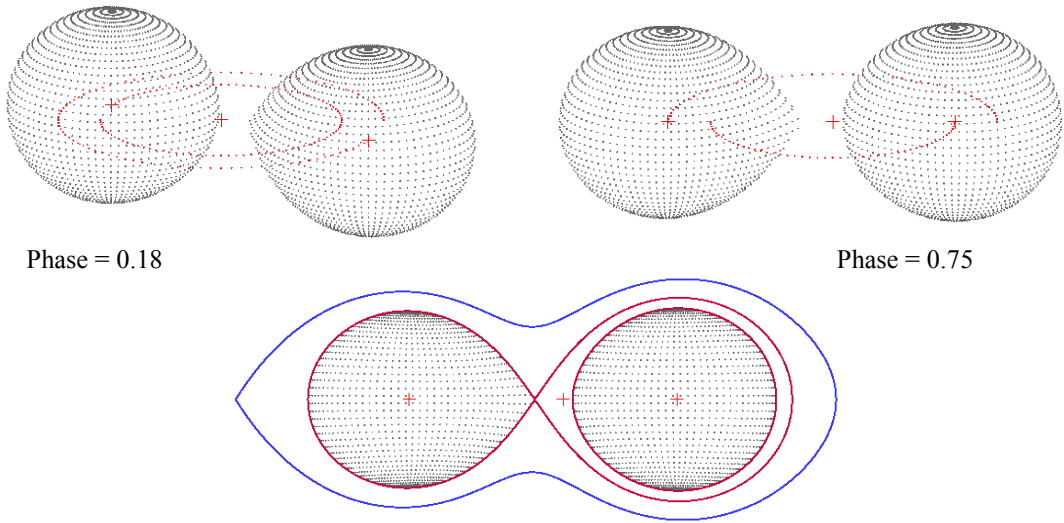

Phase = 0.18    Phase = 0.75

Figure (7): Geometric structure of the binary system GSC-03449-00680.

Table (4): Absolute physical parameters for USNO-B1.0 1091-0130715 and GSC-03449-00680.

| Element | Star Name | |
|---|---|---|
| | USNO-B1.0 1091-0130715 | GSC-03449-00680 |
| $M_1(M_\odot)$ | 1.813±0.074 | 1.331±0.054 |
| $M_2(M_\odot)$ | 1.178±0.048 | 1.414±0.058 |
| $R_1(R_\odot)$ | 1.808±0.074 | 1.358±0.055 |
| $R_2(R_\odot)$ | 1.186±0.048 | 1.334±0.055 |
| $T_1(T_\odot)$ | 1.385±0.057 | 1.116±0.046 |
| $T_2(T_\odot)$ | 0.989±0.040 | 0.712±0.029 |
| $L_1(L_\odot)$ | 12.020±0.491 | 2.956±0.121 |
| $L_2(L_\odot)$ | 1.343±0.055 | 0.954±0.039 |
| $M_{bol\_1}$ | 2.050±0.084 | 3.523±0.144 |
| $M_{bol\_2}$ | 4.430±0.181 | 7.313±0.299 |
| Sp. Type | $(A6)^1 - (G3)^2$ | $(F6)^1 - (K8)^2$ |
| D(pc) | 2239±91 | 1308±53 |

Note: [1] and [2] refer to primary and secondary components respectively

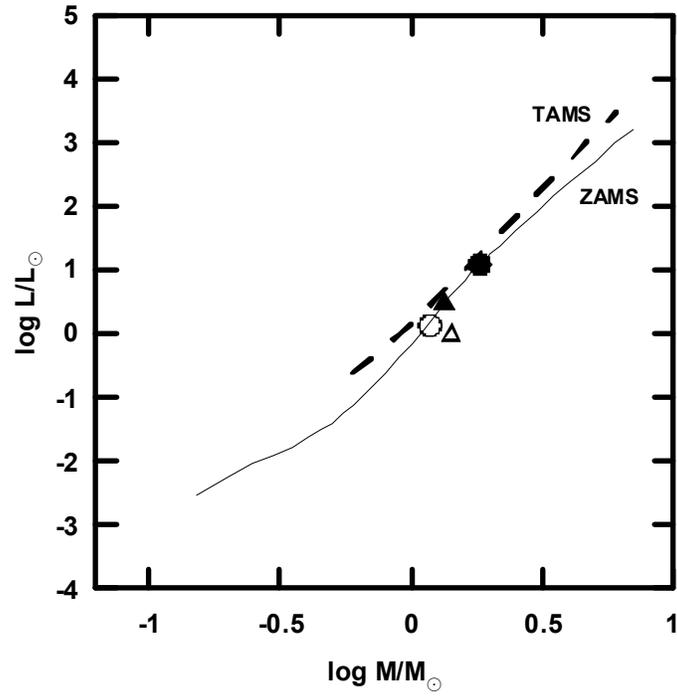

Figure (8): The position of the components of USNO-B1.0 1091-0130715 and GSC-03449-0680 on the mass–luminosity diagram. Triangles denote the components of USNO-B1.0 1091-0130715 and circles denote the components of GSC-03449-0680.

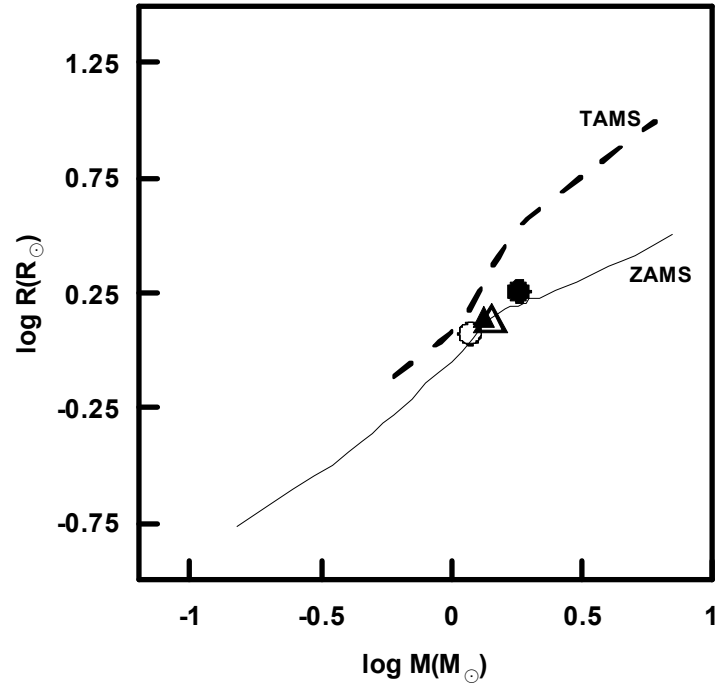

**Figure (9).** The position of the components of USNO-B1.0 1091-0130715 and GSC-03449-0680 on the mass–radius diagram. Triangles denote the components of USNO-B1.0 1091-0130715 and circles denote the components of GSC-03449-0680.

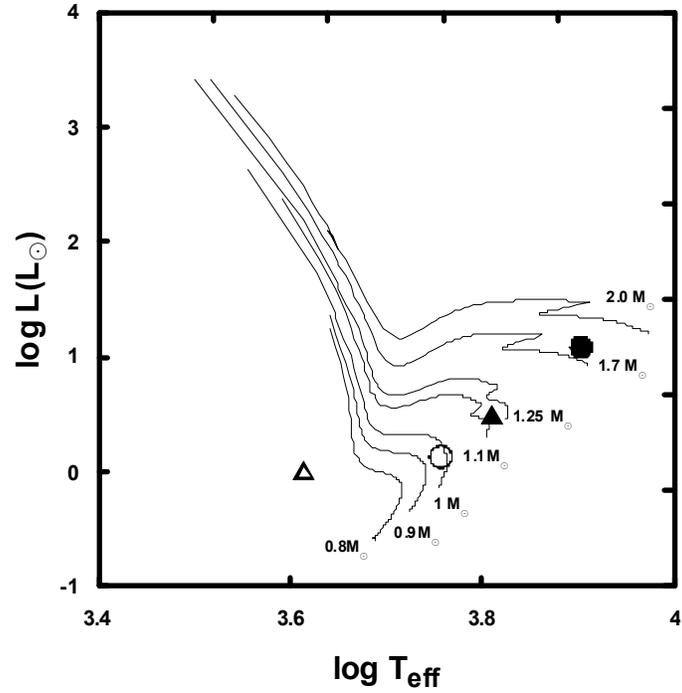

Figure (10): Position of the components of USNO-B1.0 1091-0130715 and GSC-03449-0680 on the effective temperature – luminosity diagram of Estkron et al. (2012). Triangles denote the components of USNO-B1.0 1091-0130715 and circles denote the components of GSC-03449-0680.

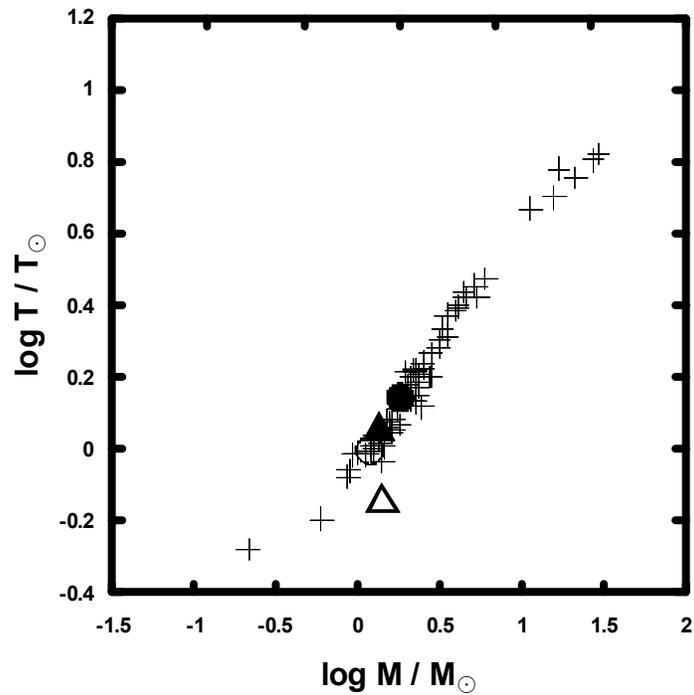

Figure (11): Position of the components of USNO-B1.0 1091-0130715 and GSC-03449-0680 on the empirical mass-Teff relation for low-intermediate mass stars by Malkov (2007). Triangles denote the components of USNO-B1.0 1091-0130715 and circles denote the components of GSC-03449-0680.